\documentstyle[preprint,tighten,aps]{revtex}

\begin{document}
\draft

\title{
The effective potential of $N$-vector models: 
a field-theoretic study to $O(\epsilon^3)$.
}
\author{Andrea Pelissetto$\,^1$ and Ettore Vicari$\,^2$ }
\address{$^1$ Dipartimento di Fisica dell'Universit\`a di Roma I
and I.N.F.N., I-00185 Roma, Italy 
}
\address{
$^2$ Dipartimento di Fisica dell'Universit\`a di Pisa and I.N.F.N., 
I-56126 Pisa, Italy \\
{\bf e-mail: \rm
{\tt Andrea.Pelissetto@roma1.infn.it},
{\tt vicari@mailbox.difi.unipi.it}
}}

\date{\today}

\maketitle

\begin{abstract}
We study the effective potential of three-dimensional O($N$) models.
In statistical physics the effective potential represents the 
free-energy density as a function of the order parameter
(Helmholtz free energy), and, therefore, it is related 
to the equation of state.
In particular, we consider its small-field expansion in the symmetric 
(high-temperature) phase, whose
coefficients are related to the zero-momentum $2j$-point renormalized 
coupling constants $g_{2j}$.  For generic values of $N$, we calculate
$g_{2j}$ to three loops in the field-theoretic approach based
on the $\epsilon$-expansion. The estimates of $g_{2j}$, or equivalently
of $r_{2j}\equiv g_{2j}/g_4^{j-1}$, are obtained by a constrained
analysis of the series that takes into account the exact results in one and 
zero dimensions. 

\medskip


\end{abstract}

\newpage


\section{Introduction}
\label{introduction}

According to the universality hypothesis, most features of 
continuous phase transitions do not depend on the microscopic details
of the systems, but only on their global properties such as the dimensionality
and the symmetry of the order parameter (see e.g.\ Ref.\ \cite{ZJbook}).
The O($N$)-symmetric universality classes describe many three-dimensional
systems characterized by short-range interactions and an $N$-component order parameter.
We mention the liquid-vapour transition in classical
fluids ($N=1$), the $\lambda$-transition in superfluid helium ($N=2$), 
the critical properties of 
isotropic ferromagnetic materials ($N=3$) and of long polymers 
($N\rightarrow 0$).
The case $N=4$ is relevant for high-energy physics:  
it should describe the critical behavior of finite-temperature QCD 
with two flavours at
the chiral-symmetry restoring phase transition~\cite{P-W-84}.
Universality implies that critical exponents, 
as well as other universal quantities, are the same for all
models belonging to the same O($N$)-symmetric class.  
Thus, the universal results obtained for a representative of a given class,
such as the O($N$)-symmetric $\phi^4$ Hamiltonian, can be used to predict the 
critical behavior of all systems in the same class.

The effective potential (Helmholtz free energy) is related to the
(Gibbs) free energy of the model.  Indeed, if $M_a\equiv\langle\phi_a\rangle$ 
is the magnetization and $H$ the magnetic field, one defines
\begin{equation}
{\cal F} (M) = M H - {1\over V} \log Z(H),
\end{equation}
where $Z(H)$ is the partition function (the dependence on the
temperature is understood).
The global minimum of the effective potential determines the value of
the order parameter which characterizes the phase of the model.  
In the high-temperature or symmetric phase the minimum is unique with
$M=0$, while in the low-temperature or broken phase
${\cal F} (M)$ presents a flat region around the 
origin~\cite{Griffiths-66}, i.e. ${\cal F} (M)$ is constant for $|M|\leq M_0$ 
where $M_0$ is the magnetization at the coexistence curve.

In the high-temperature phase the effective potential admits 
an expansion around $M=0$:
\begin{equation}
\Delta {\cal F} \equiv {\cal F} (M) - {\cal F} (0) = 
\sum_{j=1}^\infty {1\over (2j)!} a_{2j} M^{2j}.
\end{equation}
The coefficients $a_{2j}$ can be expressed in terms 
of renormalization-group invariant quantities. 
We introduce a renormalized magnetization 
\begin{equation}
\varphi^2 = {\xi(t,H=0)^2 M(t,H)^2\over \chi(t,H=0)} ,
\end{equation}
where $t$ is the reduced temperature, $\chi$ and $\xi$ are 
respectively the magnetic susceptibility and the 
second-moment correlation length
obtained from the two-point function of the order parameter $\phi$, i.e.
\begin{eqnarray}
&&\langle \phi_a(0) \phi_b(x) \rangle \equiv  \delta_{ab} G(x),\\
&&\chi = \int dx\ G(x),\qquad\qquad 
\xi^2 = {1\over 2d} {\int dx \ x^2 G(x) \over \int dx\ G(x)}.\nonumber
\end{eqnarray}
Then one may write
\begin{equation}
\Delta {\cal F}= {1\over 2} m^2\varphi^2 + 
\sum_{j=2} m^{d-j(d-2)} {1\over (2j)!} g_{2j} \varphi^{2j} .
\label{freeeng}
\end{equation}
Here $m=1/\xi$, $g_{2j}$ are functions of $t$ only, and $d$ is the space 
dimension. In field theory
$\varphi$ is the expectation value of the zero-momentum renormalized
field.  For $t\to 0$ the quantities $g_{2j}$ approach universal
constants (which we indicate with the same symbol) that represent the
zero-momentum $2j$-point renormalized coupling constants.  
A simpler parametrization of the small-field expansion of the effective potential 
can be obtained by performing a further rescaling
\begin{equation}
\varphi = {m^{(d-2)/2}\over\sqrt{g_4}} z,
\label{defzeta}
\end{equation} 
which allows us to write the free energy as
\begin{equation}
\Delta {\cal F} = {m^d\over g_4}A(z),
\label{dAZ}
\end{equation}
where
\begin{equation}
A(z) =   {1\over 2} z^2 + {1\over 4!} z^4 + 
\sum_{j=3} {1\over (2j)!} r_{2j} z^{2j},
\label{AZ}
\end{equation}
and
\begin{equation}
r_{2j} = {g_{2j}\over g_4^{j-1}} \qquad\qquad j\geq 3.
\label{r2j}
\end{equation}

The function $A(z)$ is related to the equation of state. Indeed
one can show that $z\propto t^{-\beta} M$, and
that the equation of state can be written in the form
\begin{equation}
H\propto t^{\beta\delta} {\partial A(z)\over \partial z}.
\label{eqa}
\end{equation}
The small-field expansion of the effective potential provides
the starting point for the determination 
of approximate representations of the equation of state
that are valid in the whole critical region. This requires
an analytic continuation in the 
complex $t$-plane in order to reach the coexistence curve from the 
symmetric phase~\cite{ZJbook,G-Z-97}. 
This can be achieved by using parametric 
representations \cite{Schofield_69,Schofield-etal_69,Josephson_69}, 
which implement in a rather simple way the known
analytic properties of the equation of state (Griffith's analyticity).  
This idea was successfully applied
to the Ising model, for which one can construct a systematic approximation 
scheme based on polynomial parametric representations~\cite{G-Z-97} 
and on a global stationarity condition~\cite{C-P-R-V-99},
leading to an accurate determination of
the critical equation of state and of the universal ratios of amplitudes
that can be extracted from it~\cite{G-Z-97,G-Z-98,C-P-R-V-99}. 
In view of an application  of such an approach to O($N$) models\footnote{
For $N> 1$, this approach is more difficult because of the presence of the
Goldstone singularity at the coexistence curve,
which must be somehow taken into account
by the approximate parametric representations considered.} with $N> 1$,
we decided to improve the estimates of the coefficients 
$r_{2j}$ appearing in Eq. (\ref{AZ}).
It is worth mentioning that a better determination  of the equation of state,
and therefore of the universal ratios of amplitudes such as
the ratio of the specific heat amplitudes $A^+/A^-$, is 
particularly important in the case $N=2$, which describes the 
$\lambda$-transition in ${}^4$He.  
A recent Space Shuttle esperiment~\cite{Lipa-etal-96} made a
very precise measurement of the heat capacity of liquid 
helium to within 2 nK from
the $\lambda$-transition, obtaining extremely accurate estimates of
the exponent $\alpha$ and of the ratio $A^+/A^-$. These results
represent a challenge for theorists, who, until now, have not 
been able to compute universal quantities
at the same level of accuracy (see e.g. Refs.~\cite{G-Z-98,S-L-D-99,H-T-99,C-P-R-V-99-b} for recent
theoretical estimates of $\alpha$ and $A^+/A^-$).

Several approaches can be employed to investigate the O($N$)-vector models, such as
field-theoretical methods starting  from the $\phi^4$ formulation of the theory
\begin{equation}
{\cal H}= \int d^dx\left[ {1\over 2}\partial_\mu \phi(x)\partial_\mu \phi(x)
+ {1\over 2}r \phi^2 + {1\over 4!}g_0(\phi^2)^2\right],
\label{Hphi4}
\end{equation}
lattice techniques performing  high- and low-temperature expansions, 
Monte Carlo simulations, etc...
In order to study the small-field expansion of the effective potential, 
we consider the 
field-theoretic approach based on the 
$\epsilon\equiv 4-d$ expansion~\cite{W-F-72}.
We extend the series of $r_{2j}$ to $O(\epsilon^3)$
(corresponding to a three-loop calculation) for generic values of $N$,
and obtain new estimates from their analysis.
Earlier estimates of $r_{2j}$ based on the $\epsilon$-expansion~\cite{P-V-98-b}
were obtained using $O(\epsilon^2)$ series for generic values of $N$
and $O(\epsilon^3)$ series for the Ising model, which
were derived from the scaling equation of state known 
to $O(\epsilon^2)$ for generic values of $N$~\cite{B-W-W-72}
and to $O(\epsilon^3)$ for the Ising model~\cite{W-Z-74,N-A-85}.

Since the  $\epsilon$-expansion is asymptotic, one needs to perform 
a resummation of the
series in order to obtain reliable estimates. This can be efficiently done
by exploiting its Borel summability and the knowledge
of the large-order behavior~\cite{L-Z-77}. Moreover, one may exploit
exact results for low-dimensional models and perform 
constrained analyses of the $\epsilon$-series. The basic assumption 
is that the zero-momentum $2j$-point renormalized couplings $g_{2j}$, and therefore the ratios $r_{2j}$,
are analytic and quite smooth in the domain
$4>d>0$ (thus $0 < \epsilon < 4$). This can be verified in the large-$N$ limit~\cite{P-V-98,P-V-98-b}.
One may then perform a polynomial interpolation among 
the values of $d$ where the constants $r_{2j}$ are known, 
and then analyze the series of the difference.
As we shall see, the analysis of the $O(\epsilon^3)$  series of $r_{2j}$
leads to a substantial improvement of the estimates of the first few $r_{2j}$
with respect to earlier results obtained from their 
$O(\epsilon^2)$ series~\cite{P-V-98-b}.
As a by-product of our analysis we also obtain new estimates 
for the first few $r_{2j}$ in the  two-dimensional O($N$) models. 

The zero-momentum
four-point coupling $g\equiv g_4$ plays an important role in the field-theoretic perturbative expansion
at fixed dimension~\cite{Parisi-73}, which provides an accurate description of the critical region 
in the symmetric phase.
In this approach, any universal quantity is obtained from a series 
in powers of $g$ ($g$-expansion), which is then resummed and evaluated at 
the fixed-point value of $g$, $g^*$ (see e.g. Refs. \cite{ZJbook}).
Accurate estimates of $g^*$ have been obtained by 
calculating the zero of the Callan-Symanzik $\beta$-function associated
to $g$ (see e.g. Refs.~\cite{G-Z-98,A-S-95,Nickel-91,L-Z-77,B-N-G-M-77,ZJbook}).
These results have been substantially confirmed by computations using 
different approaches, such as $\epsilon$-expansion~\cite{P-V-98},
high-temperature expansion (see, e.g., 
Refs.~\cite{C-P-R-V-99,B-C-98} and references therein), 
Monte Carlo simulations~\cite{Tsypin-94,B-K-96,Kim-99}, etc...
In this paper we reconsider the determination of $g^*$ from 
its $\epsilon$-expansion.
In Ref.~\cite{P-V-98} we calculated it to $O(\epsilon^4)$, 
but unfortunately the series published there  contains a  numerical 
mistake\footnote{The numerical expressions for the Feynman graphs 
appearing in Ref.~\cite{P-V-98} have been checked in Ref. 
\cite{Chung-Chung}. All numerical estimates are in agreement, 
except that of the constant $H$. The correct value is given in the Appendix.}.
For this reason, we report here the correct series and the results of the new
analysis.
We anticipate that the changes with respect to the estimates reported in Ref.~\cite{P-V-98}
are very small.

The field-theoretic method based on the $g$-expansion at fixed dimension has
been recently considered in the calculation of 
the zero-momentum couplings $g_{2j}$ with $j > 2$: for the Ising model 
five-loop series~\cite{G-Z-98,G-Z-97,B-B-M-N-87} are available, while 
for generic values of $N$, $g_6$ and $g_8$ have been determined to
four and three loops respectively~\cite{S-O-U-K-99,S-O-98,S-O-U-97}. 
The effective potential has also been studied by
approximately solving the exact renormalization-group equations,
providing some estimates of the 
coefficients $g_{2j}$~\cite{T-W-94}. For the Ising
model accurate estimates have been obtained from the analysis 
of lattice high-temperature
expansions, see e.g. Refs.~\cite{C-P-R-V-99,B-C-97,Z-L-F-96}.
For $N> 1$, which is the case considered in this paper,
only the high-temperature of $g_6$ has been computed~\cite{Reisz-95}:
however, the results of its analysis are rather imprecise.

The paper is organized as follows.
In Sec.~\ref{sec2} we present the $O(\epsilon^3)$ 
series of $r_{2j}$ that we have calculated. 
In Sec.~\ref{sec3} we give the results of the analyses of the series 
of $r_{2j}$,
which are then compared with the available estimates obtained in 
other approaches. The appendix is dedicated to the calculation
of the three-loop integrals involved in the computation of the 
zero-momentum $n$-point irreducible functions.

\section{Expansion of $\lowercase{r}_{2\lowercase{j}}$ to $O(\epsilon^3)$.}
\label{sec2}

In the framework of the $\epsilon$-expansion we have calculated, to three loops, the 
one-particle irreducible correlation functions at zero momentum
\begin{equation}
\Gamma_{2j}\equiv \Gamma^{(2j)}_{\alpha_1\alpha_1...\alpha_j\alpha_j}
(0,...,0).
\end{equation}
The number of diagrams one has to evaluate to compute $\Gamma_{2j}$
increases with increasing $j$. For example, 
the three-loop one-vertex irreducible diagrams necessary to compute 
$\Gamma_6$, $\Gamma_8$, $\Gamma_{10}$ are
16, 36, 64 respectively.
In our calculation we employed a symbolic manipulation package developed in 
{\sc Mathematica}. 
It generates the diagrams using the algorithm described in Ref.~\cite{Heap-66},
performs the necessary index contractions to determine 
the $N$ dependence of  each diagram, and compute the corresponding integral
according to the procedure described in App.~\ref{appa}.

The coefficients $r_{2j}$ of the expansion of $A(z)$ in powers of $z$ 
can be written in terms of 
$\Gamma_{2j}$  as 
\begin{equation}
r_{2j} \equiv {g_{2j}\over g^{j-1}}= {(2j)!\over 2^j3^{j-1}j!}
{(N+2)^{j-2}\over  \prod_{i=2}^{j-1} (N+2i)}
{\Gamma_{2j} \Gamma_2^{j-2}\over \Gamma_4^{j-1}}.
\end{equation}
From the three-loop expansion of $\Gamma_{2j}$, one can derive the 
series of $r_{2j}$ to $O(\epsilon^3)$.
In the following we report the series of the first few $r_{2j}$, 
i.e. $r_6$, $r_8$ and $r_{10}$, that
we will analyze in the following section. 
Writing
\begin{equation}
r_{2j} = \sum_{i=1} r_{2j,i}\,\epsilon^i ,
\end{equation}
we find
\begin{eqnarray}
r_{6,1} = && {\frac{5(26+N)}{6(8+N)}}      \\ 
r_{6,2} = &&     -{\frac{98 + 33\,N + 4\,{N^2}}{{{\left( 8 + N \right) }^3}}}+ 
    {\frac{40\,\lambda\,\left( -8 + 7\,N + {N^2} \right)}{3\,
      {{\left( 8 + N \right) }^3}}} \nonumber \\
r_{6,3} = &&
-   {\frac{5\,\left( 17264 + 9968\,N + 2574\,{N^2} + 319\,{N^3} + 7\,{N^4}
          \right) }{6\,{{\left( 8 + N \right) }^5}}} \nonumber \\ 
&&+   {\frac{5\,\lambda\,\left( -2176 - 172\,N + 152\,{N^2} + 9\,{N^3} \right) }
     {3\,{{\left( 8 + N \right) }^4}}}  
+  {\frac{20\,(Q_1+\gamma_E\,\lambda)\,\left( N -1\right) }
     {3\,{{\left( 8 + N \right) }^2}}}   \nonumber \\ 
&&
+   {\frac{640\,\left( N-1 \right) \,Q_2}
     {{{\left( 8 + N \right) }^3}}} 
+   {\frac{20\,\left( 682 + 49\,N - 2\,{N^2} \right) \,\zeta(3)}
     {{{\left( 8 + N \right) }^4}}} 
\nonumber
\end{eqnarray}
\begin{eqnarray}
r_{8,1} = && -{\frac{35(80+N)}{18(8+N)}}    \\ 
r_{8,2} = &&   {\frac{35\,\left( 31904 + 7610\,N + 578\,{N^2} + 3\,{N^3} \right) }{54\,
      {{\left( 8 + N \right) }^3}}} -
 {\frac{5600\,\lambda\,\left( -8 + 7\,N + {N^2} \right)  }{27\,
      {{\left( 8 + N \right) }^3}}} \nonumber \\
r_{8,3} = &&
   {\frac{35\,\left( -259712 - 112232\,N - 16204\,{N^2} - 422\,{N^3} +
         13\,{N^4} \right) }{18\,{{\left( 8 + N \right) }^5}}} 
\nonumber \\
&&
+   {\frac{35\,\lambda\,\left( 105472 + 72528\,N - 384\,{N^2} - 
         469\,{N^3} \right) }{81\,{{\left( 8 + N \right) }^4}}} 
+  {\frac{2800\,(Q_1+\gamma_E\,\lambda)\,\left( 1 - N \right) }
     {27\,{{\left( 8 + N \right) }^2}}}  \nonumber\\
&&
+   {\frac{85120\,\left( 1 - N \right) \,Q_2}
     {9\,{{\left( 8 + N \right) }^3}}} 
+   {\frac{70\,\left( -29824 - 3010\,N + 29\,{N^2} \right) \,
       \zeta(3)}{9\,{{\left( 8 + N \right) }^4}}}  \nonumber
\end{eqnarray}
\begin{eqnarray}
r_{10,1} = && {\frac{35(242+N)}{3(8+N)}}      \\ 
r_{10,2} = &&  - {\frac{35\,\left( 2083280 + 453428\,N + 28580\,{N^2} + 63\,{N^3}\right)}{108\,
      {{\left( 8 + N \right) }^3}}} +
{\frac{162400\,\lambda\,\left( -8 + 7\,N + {N^2} \right) }{27\,
      {{\left( 8 + N \right) }^3}}} \nonumber \\
r_{10,3} = &&
   {\frac{35\,\left( 157284800 + 62464976\,N + 8716080\,{N^2} + 
         388468\,{N^3} - 110\,{N^4} + 27\,{N^5} \right) }{108\,
       {{\left( 8 + N \right) }^5}}}  \nonumber \\
&&+   {\frac{140\,\lambda\,\left( -739808 - 822816\,N - 35058\,{N^2} + 
         3359\,{N^3} \right) }{81\,{{\left( 8 + N \right) }^4}}}  \nonumber \\
&&
+  {\frac{81200\,(Q_1+\gamma_E\,\lambda)\,\left( N-1 \right) }
     {27\,{{\left( 8 + N \right) }^2}}} 
+   {\frac{1433600\,\left( N-1 \right) \,Q_2}
     {9\,{{\left( 8 + N \right) }^3}}} \nonumber \\
&&
+   {\frac{140\,\left( 463924 + 65932\,N + 1585\,{N^2} \right) \,
       \zeta(3)}{9\,{{\left( 8 + N \right) }^4}}} \nonumber
\end{eqnarray}
where
\begin{eqnarray}
\lambda &=&\hphantom{-} 1.171 953 619 344 729 445... \\
Q_1     &=&          -  2.695 258 053 506 736 953... \nonumber \\
Q_2     &=&\hphantom{-} 0.400 685 634 386 531 428... \nonumber 
\end{eqnarray}
For $N=1$ the above expressions reproduce the $O(\epsilon^3)$ series
that can be derived  from the $O(\epsilon^3)$ equation of state
calculated in Refs.~\cite{W-Z-74,N-A-85}.
We could also have computed the $O(\epsilon^3)$ series of 
$r_{2j}$ for some additional $j>5$. Since, 
as we shall see, with increasing $j$, longer and longer series are necessary
to obtain acceptable three-dimensional estimates from their analysis, we 
decided not to go beyond $r_{10}$.

We now report the series of the fixed-point values of the
zero-momentum four-point coupling $g\equiv g_4$,
which corrects that given in Ref.~\cite{P-V-98}, 
which was plagued by a numerical 
mistake in one of the three-loop integrals.
In the framework of the $\epsilon$-expansion, we found convenient
to consider the rescaled coupling $\bar{g}$, defined by 
\begin{equation}
\bar{g} \equiv 
   {1\over 2 (4\pi)^{d/2}} {(N+8)\over3} \Gamma\left(2 - {d\over2}\right)g\; ,
\label{gbar}
\end{equation}
and expand it in powers of $\epsilon$.
Due to the $O(\epsilon^{-1})$
factor multiplying $g$ in the definition of $\bar{g}$, 
 the $O(\epsilon^4)$ of $g^*$
corresponds to the $O(\epsilon^3)$ of $\bar{g}^*$.
The $\epsilon$-expansion of $\bar{g}^*$ to $O(\epsilon^3)$
is given by
\begin{equation}
   \bar{g}^{*}(\epsilon) = \sum_{k=0} \bar{g}_k \epsilon^k\; ,
\label{gepsseries}
\end{equation}
with
\begin{eqnarray}
\bar{g}_0 =&& 1 \label{gbars} \\
\bar{g}_1 =&& {3(14+3N)\over (8 + N)^2} \nonumber \\
\bar{g}_2 =&& {\frac{1224 + 520N + 58N^2 -2N^3}{(8+N)^4}}
    - {\frac{ 12(22+5N)\zeta(3)}{(8+N)^3}}
 - {\frac{\lambda (62+13N)}{3(8+N)^2}} \nonumber \\
\bar{g}_3 =&& 
   {\frac{341312 + 225312\,N + 57572\,{{N}^2} + 
       5404\,{{N}^3} - 99\,{{N}^4} + 
       4\,{{N}^5}}{8\,{{\left( 8 + N \right) }^6}}}
     - {\frac{\left( 22 + 5\,N \right) \,{{\pi }^4}}
     {15\,{{\left( 8 + N \right) }^3}}} \nonumber \\
&&
+   {\frac{2\,\left( -3880 - 772\,N + 431\,{{N}^2} + 
         90\,{{N}^3} \right) \,\zeta(3)}{{{\left(
           8 + N \right) }^5}}} + 
   {\frac{40\,\left( 186 + 55\,N + 2\,{{N}^2}
         \right) \,\zeta(5)}{{{\left( 8 + N \right) }^
        4}}}\nonumber \\
&&  -{\frac{\lambda (6500 + 2700 N + 327 N^2  + 4 N^3 )}{2(8+N)^4}}
-   {\frac{(Q_1+\gamma_E\,\lambda)\,
       \left( 62 + 13\,N \right) }{6\,
       {{\left( 8 + N \right) }^2}}} \nonumber \\ 
&&-    {\frac{8\,\left( 62 + 19\,N \right) \,Q_2}
     {{{\left( 8 + N \right) }^3}}} -
  {\frac{8\,H\,\left( 22 + 5\,N \right) }
     {{{\left( 8 + N \right) }^3}}} \; ,
\nonumber
\end{eqnarray}
where
\begin{equation}
H      =         -  2.155 952 487 340 794 361...
\end{equation}

Finally we report the exact results in $d=1$ and $d=0$ that we will use in our 
constrained analyses of the $\epsilon$-series.
In $d=1$ we have~\cite{P-V-98-b}
\begin{eqnarray}
r_6 =&& 5 - \frac{5 N (N-1)^2 (8 N + 7)}{(N+1) (N+4) (4 N - 1)^2}\; ,
\label{d1res} \\
r_8=&& {175\over3} - 
    \frac{35 N (N-1)^2 (256 N^3 + 3037 N^2 + 1705 N - 588}{
         3 (N+1) (N+4) (N+6) (4 N - 1)^3}\; ,
 \nonumber \\
r_{10}= &&
1225 - 175 N (N - 1)^2 (N+1)^{-2} (N+3)^{-1}(N+4)^{-2}(N+6)^{-1}(N+8)^{-1}(4N-1)^{-4}\times\nonumber\\
&& (149184 - 886968N - 690826N^2 + 4219985N^3 + 6283975N^4 + 
        2913758N^5 \nonumber \\
&&+ 552223N^6 + 44405N^7 + 1664N^8)\; ,
\nonumber
\end{eqnarray}
for $N\geq 1$, and
\begin{eqnarray}
r_6 &=& 5, \\
r_8&=& {175\over 3},\nonumber \\
r_{10}&=&  1225, \nonumber 
\end{eqnarray}
for $N\leq 1$. In $d=0$ we have
\begin{eqnarray}
r_6 &=& {10(N+8)\over 3(N+4)},
\label{d0res} \\
r_8&=&{70(N^2+14N+120)\over 3(N+4)(N+6)},
 \nonumber \\
r_{10}&=& 
         {280 (10752 + 3136 N + 256 N^2  + 30 N^3  + N^4 ) \over
                  (N+4)^2 (N+6) (N+8)}, \nonumber 
\end{eqnarray}
for $N\geq 1 $. 
It is not clear how to determine the value of $r_{2j}$ for $N=0$.
Similarly to the $d=1$ case, one may conjecture
that their values are independent of $N$ for $N\leq 1$, 
and therefore equal to those for $N=1$.

\section{Analyses of the series.}
\label{sec3}

Since the $\epsilon$-expansion is asymptotic,
it requires a resummation 
to get estimates for $d=3$, i.e. $\epsilon=1$,
which is usually performed assuming its Borel summability. 
The analysis of the series in powers of $\epsilon$ can be performed
by using the method proposed in Ref.~\cite{L-Z-77}, which
is based on the knowledge of the large-order behaviour of the series.
It is indeed known  that the $n$-th coefficient of the series 
behaves as $\sim (-a)^n \Gamma(n + b_0 + 1)$ for large $n$.
The constant $a$, which characterizes the singularity of the 
Borel transform,  does not depend 
on the specific quantity; it is given by~\cite{Lipatov-77,B-L-Z-77} 
$a = 3/(N + 8)$.
The coefficient $b_0$ depends instead on the quantity at hand.

Consider a generic quantity $R$ whose $\epsilon$-expansion is 
\begin{equation}
R(\epsilon)=\sum_{k=0} R_k \epsilon^k.
\end{equation}
According to Ref.~\cite{L-Z-77},
one generates new series $R_p(\alpha,b;\epsilon)$ according to
\begin{equation}
R_p(\alpha,b;\epsilon) = \sum_{k=0}^p 
    B_k(\alpha,b) 
  \int^\infty_0 dt\ t^b\
  e^{-t} {u(\epsilon t)^k \over \left[1 - u(\epsilon t) \right]^\alpha}  \;,
\label{RBorel}
\end{equation}
where 
\begin{equation}
   u(x) = { \sqrt{1 + a x} - 1\over \sqrt{1 + a x} + 1}.
\end{equation}
The coefficients $B_k(\alpha,b)$ are determined by the requirement
that the expansion in $\epsilon$ of $R_p(\alpha,b;\epsilon)$
coincides with the original series. For each $\alpha$, $b$ and $p$
an estimate of $R$ is simply given by $R_p(\alpha,b;\epsilon=1)$.

We follow Ref.~\cite{P-V-98} to derive the estimates and their uncertainty.
We determine an integer value of $b$, $b_{\rm opt}$, such that
\begin{equation} 
R_K(\alpha,b_{\rm opt};\epsilon=1)
    \approx R_{K-1}(\alpha,b_{\rm opt};\epsilon=1),
\end{equation}
for $\alpha < 1$, where $K$ is the highest order of the known 
terms in the series.
In a somewhat arbitrary way 
we consider as our final estimate the average of 
$R_K(\alpha,b;\epsilon=1)$ with 
$-1 < \alpha \le 1/2$ and $-2 + b_{\rm opt} \le b \le 2 + b_{\rm opt}$.
The error we report is the variance of the values of
$R_K(\alpha,b;\epsilon=1)$
with $-1 < \alpha \le 1/2$ and 
$\lfloor b_{\rm opt}/3 - 1\rfloor \le b \le \lceil  4 b_{\rm opt}/3 + 1\rceil$.
This procedure was already discussed and tested in 
Refs.~\cite{P-V-98,P-V-98-b}.
It seems to provide reasonable estimates and error bars, at least
when a sufficiently large number of terms is known.
Nonetheless, the method is {\em ad hoc}, and its reliability 
may depend on the quantity at hand.
It is therefore important to check the accuracy of the error estimates 
for each quantity considered, 
for example by comparing, when it is possible, results
obtained from different series representing the same quantity.

It has been noted that, 
if one assumes a sufficiently large analytic domain in $d$ for the 
quantity at hand,
the estimates from its $\epsilon$-expansion can be improved using 
its known value
in lower dimensions~\cite{DesClo-81,L-Z-87,C-P-R-V-98,P-V-98,P-V-98-b,P-V-99}.
One can check explicitly in the large-$N$ limit
that the zero-momentum couplings  $g_{2j}$ are analytic  for $0<d<4$. One may then 
conjecture that the analytic properties of $g_{2j}$ emerged in the large-$N$ limit
are also valid for finite fixed values of $N$. 
This point was already discussed in Refs.~\cite{P-V-98,P-V-98-b}.

Suppose that the 
values of the generic quantity $R$ are known for a set of dimensions 
$\epsilon_1$,...,$\epsilon_k$. In this case 
one may use as zeroth order approximation the value for $\epsilon=1$ of the polynomial 
interpolation through $\epsilon=0$, $\epsilon_1$,...,$\epsilon_k$ and 
then use the series 
in $\epsilon$ to compute the deviations. 
More precisely, let us suppose that exact values  
$R_{\rm ex}(\epsilon_1)$, $\ldots$, $R_{\rm ex}(\epsilon_k)$ are known 
for the set of dimensions $\epsilon_1$, $\ldots$, $\epsilon_k$, 
$k \ge 2$. Then define 
\begin{equation}
Q(\epsilon) = \sum_{i=1}^k \left[
    {R_{\rm ex}(\epsilon_i) \over (\epsilon - \epsilon_i)}
    \prod_{j=1,j\not=i}^k (\epsilon_i - \epsilon_j)^{-1} \right]\; ,
\end{equation}
and 
\begin{equation}
S(\epsilon) = 
   {R(\epsilon) \over \prod_{i=1}^k (\epsilon - \epsilon_i)} - 
   Q(\epsilon) ,
\label{ses}
\end{equation}
and finally
\begin{equation}
   R_{\rm imp}(\epsilon) = \left[ Q(\epsilon) + 
      S(\epsilon) \right]
      \prod_{i=1}^k (\epsilon - \epsilon_i)  .
\label{serieconstrained}
\end{equation}
The resummation procedure is applied to $S(\epsilon)$
and the final estimate is obtained by computing $R_{\rm imp}(\epsilon=1)$.
If the polynomial
interpolation in $d$ is a good approximation, 
one should find that the $\epsilon$-series which gives the 
deviations has smaller coefficients than the original one. 
Consequently one expects that also the errors in the resummation are reduced.
In the cases considered, we find that, as expected, the coefficients
of the series $S(\epsilon)$ decrease in size with $k$, 
the number of exact values that are used to constrain the series.

To begin with, we present the results of the analyses of the series for 
four-point coupling $\bar{g}^*$, cf. Eq.~(\ref{gbars}). 
Table~\ref{epsilon3d} shows the three-dimensional results
for various constrained analyses, using the exact results in $d=1$ and 
$d=0$ and the accurate estimates for two-dimensional models~\cite{P-V-98}.
Note that, for each $N$, the results of the various constrained analyses  
are consistent with each other. 
The two-dimensional estimates used in the constrained analyses are those 
already considered in Ref.~\cite{P-V-98}, except for $N=3$.
In this case 
we used the recent  two-dimensional estimate 
$\bar{g}^*=1.7778(45)$ obtained from
a form-factor bootstrap approach~\cite{B-N-N-P-S-W-99}, which should be 
more reliable than the high-temperature result 
$\bar{g}^*=1.724(9)$ considered in Ref.~\cite{P-V-98}.
For comparison, we note that for the two-dimensional O(3) model 
our new estimate from the $\epsilon$-expansion is 
$\bar{g}^*=1.75(3)$. In Table~\ref{epsilon2d} we present the 
two-dimensional results obtained
by analyses constrained in $d=1$ and $d=0$.

Tables~\ref{epsilon3d} and \ref{epsilon2d}  
supersede the corresponding ones, i.e. Tables 1 and 2, of Ref.~\cite{P-V-98}.
The changes are small, 
so that the discussion presented there remains valid.
In Table~\ref{summarygr} we report the additional recent estimates which
appeared after Ref.~\cite{P-V-98}.
These results have been obtained from:
(i) a reanalysis of the six-loop $\beta$-function in the framework
of the fixed-dimension $g$-expansion~\cite{G-Z-98};
(ii) the analyses of the high-temperature expansions (HT) 
of O($N$) $\sigma$-models on the cubic and bcc lattices~\cite{B-C-98};
(iii) high-temperature expansions of improved Hamiltonians (IHT)
for which the leading scaling corrections are suppressed~\cite{C-P-R-V-99,C-P-R-V-99-b,C-P-R-V-00}.
The agreement among the various estimates is globally good.  Let us only note
that for $N=3$ the error we obtained from our analysis of the $\epsilon$-series
seems to be underestimated.
A more complete list of references presenting estimates of 
$\bar{g}^*$ can be found in Refs.~\cite{P-V-98,G-Z-98,C-P-R-V-99}.

Since the series of $r_{2j}$ begins with 
a term of order $\epsilon$, we analyzed the $O(\epsilon^2)$ series of the
quantity $r_{2j}/\epsilon$.
The constrained analyses were performed using the exact results 
of $r_{2j}$ for $d=1$ and $d=0$.
We mention that in the case of the Ising model ($N=1$)
one may also use the precise two-dimensional estimates obtained from 
the analyses of the available high-temperature series for the 
Ising model on the square and triangular lattice~\cite{K-Y-T-77,McKenzie-79}.  
These results were exploited 
in the analysis of the $O(\epsilon^3)$ series of the Ising model presented
in Ref.~\cite{P-V-98-b}, allowing us to further 
improve the estimates of $r_{2j}$.  A discussion of the large-$N$ 
limit of $r_{2j}$ can be found in Ref.~\cite{P-V-98-b}.

Tables~\ref{r6d3} and \ref{r6d2} show respectively our three- and 
two-dimensional results
for $r_6$. There we report the  estimates obtained from an 
unconstrained analysis and from analyses constrained in
various dimensions. Like the case of $g^*$, the results of the various analyses
are consistent with each other and the error decreases
when additional lower dimensional values are used to constrain the  analysis, 
supporting the estimate and the error
obtained by the $d=0,1$ constrained analyses.

Tables~\ref{r8d3} and \ref{r8d2} present respectively our 
three- and two-dimensional results
for $r_8$. 
In this case, the results of the different analyses are not in complete 
agreement, indicating a possible underestimate of the uncertainty.
Therefore, we believe that the final estimates of $r_8$ obtained from 
the $d=0,1$ constrained analyses have an error  larger than what is given 
by our algorithmic procedure. In Table \ref{summaryd3},
where we report our final estimates,  we multiply the 
errors by a factor of two: this quoted uncertainty should be more 
realistic.

In the case of  $r_{10}$, the analyses, even those constrained, 
of its $O(\epsilon^3)$ series
do not provide satisfactory estimates, but give only 
an order of magnitude. 
This may be explained by looking at the
coefficients of the series of $r_{2j}$: they increase rapidly with $j$, 
and, for $r_{10}$, 
they are already much larger than its final three-dimensional estimate.
We just mention that the $d=0,1$ constrained
analyses give $r_{10} = 29(34)$ for $N=2$, 
$r_{10} = 16(24)$ for $N=3$, $r_{10} = 9(17)$ for $N=4$, where the 
reported errors
are those obtained from our algorithmic procedure. 

Let us compare our results with the available estimates from other approaches. 
For $N\neq 1$  there are not many published results: we are only aware
of the estimates of $g_6$ and $g_8$ presented in 
Refs.~\cite{S-O-U-K-99,Reisz-95,T-W-94}
(from which we can derive estimates of $r_6$ and $r_8$ using their 
results for $g_4$).  Table~\ref{summaryd3} presents a 
summary  of the available estimates of $r_6$, $r_8$ and $r_{10}$
for several values of $N$. 
For the sake of completeness, we report also the results for $N=1$
although the three-loop series were already known and no new results we have
obtained in this work 
(a more complete list of results for the Ising model can be found
in Ref.~\cite{C-P-R-V-99}).

In Ref.~\cite{S-O-U-K-99} $g_6$ and $g_8$ were  estimated from a Pad\'e-Borel
resummation of their $d=3$ $g$-expansion, calculated to 
four and three loops respectively. The authors of Ref.~\cite{S-O-U-K-99} argue
that the uncertainty on their estimates of $g_6$ is approximately 0.3\%,
while they consider their values for $g_8$ much less accurate.
These results are in good agreement with ours, 
especially those for $r_6$. We have also redone
the analysis of the four-loop series of $r_6$ calculated 
in Ref.~\cite{S-O-U-K-99} using the same method employed here 
for the $\epsilon$-expansion
(the results are reported in  Table~\ref{summaryd3}). 
In Table~\ref{summaryd3} we also show some results for the $N=2$ model
obtained from the analysis of the 20th-order 
high-temperature expansion of an improved
lattice Hamiltonian~\cite{C-P-R-V-00}: 
$r_6$ is in good agreement with the $\epsilon$-
and $g$-expansion, while for $r_8$ IHT seems to give the most precise estimate.
We also mention that Ref.~\cite{T-W-94} uses a renormalization-group 
approach (ERG) in which
the exact RG equation is approximately solved
(no estimates of the errors are presented there). 
Concerning $r_{10}$, 
the result for $N=2$ can be compared with the estimate coming from
IHT: $r_{10}=-13(7)$~\cite{C-P-R-V-00}.

As already mentioned in the introduction, the estimates of the first few 
$r_{2j}$ that we have presented in this work
may be very useful for the determination of the whole critical equation of state
in O($N$) models with $N>1$. Work is indeed in progress~\cite{C-P-R-V-00}.
This approach was already successfully applied to the Ising model. In the 
cases $N>1$ the presence of the Goldstone singularity requires 
a more careful treatement.

\acknowledgements
We thank Jin-Mo Chung for useful correspondence.

\appendix

\section{Three-loop diagrams.}
\label{appa}

In the three-loop calculations of the zero-momentum irreducible $2j$-point functions one deals
with four families of integrals:

\begin{eqnarray}
B_2(n_1,n_2,n_3) &\equiv& 
  \int {d^dk_1\over (2\pi)^d} {d^dk_2\over (2\pi)^d} \,
   \Delta(k_1)^{n_1} \Delta(k_2)^{n_2} \Delta(k_1+k_2)^{n_3},
\label{defB2}
\\
B_3(n_1,n_2,n_3,n_4) &\equiv&
  \int {d^dk_1\over (2\pi)^d} {d^dk_2\over (2\pi)^d} 
       {d^dk_3\over (2\pi)^d} 
   \Delta(k_1)^{n_1} \Delta(k_2)^{n_2} \Delta(k_3)^{n_3} 
   \Delta(k_1+k_2+k_3)^{n_4},
\nonumber \\ [-2mm]
\label{defB3}
{} \\
G(n_1;n_2,n_3;n_4,n_5) &\equiv&
  \int {d^dk_1\over (2\pi)^d} {d^dk_2\over (2\pi)^d} 
       {d^dk_3\over (2\pi)^d}
   \Delta(k_1)^{n_1} \Delta(k_2)^{n_2} \Delta(k_3)^{n_4} 
\nonumber \\[2mm]
&& \qquad 
   \times\,  \Delta(k_1+k_2)^{n_3} \Delta(k_1+k_3)^{n_5},
\label{defG}
\\[2mm]
M(n_1,n_2,n_3;n_4,n_5,n_6) &\equiv&
   \int {d^dk_1\over (2\pi)^d} {d^dk_2\over (2\pi)^d}  
       {d^dk_3\over (2\pi)^d}
   \Delta(k_1)^{n_1} \Delta(k_1+k_2)^{n_2} \Delta(k_1+k_3)^{n_3} 
\nonumber \\[2mm]
&& \qquad \times\, 
   \Delta(k_2-k_3)^{n_4} \Delta(k_3)^{n_5} \Delta(k_2)^{n_6},
\label{defM}
\end{eqnarray}
where $\Delta(k)$ is the massive propagator, 
\begin{equation}
\Delta(k) \equiv {1\over k^2 + 1}.
\end{equation}
In order to compute these integrals, we have used an algorithm 
that expresses each quantity in terms of four basic integrals:
$B_2(1,1,1)$, $B_3(1,1,1,1)$, $G(1,1,1,1,1)$ and $M(1,1,1,1,1,1)$.
The algorithm is based on the integration-by-parts 
technique \cite{Chetyrkin-Tkachov_81,Tkachov_81,Avdeev_96}. 
The reduction of the integrals of type $B_3$ is accomplished using 
the method presented in Ref. \cite{Broadhurst_92}, noting that 
these integrals are related to $B_N(0,0,n_1,n_2,n_3,n_4)$, as defined 
in Ref. \cite{Broadhurst_92}. The integral $B_3(1,1,1,1)$ can be 
expressed in terms of the constant $B_4$ defined in \cite{Broadhurst_92}
since
\begin{eqnarray}
B_4 &=& - {4\over \epsilon^5} (1 - 3 \epsilon) - 
  {14\over 3 \epsilon^4} \,
  {\Gamma(1-\epsilon/2) \Gamma(1 + \epsilon)^2 \Gamma(1 + 3\epsilon/2) \over
   \Gamma(1+\epsilon/2)^2 \Gamma(1 + 2\epsilon)} 
\nonumber \\
&& + {(1-\epsilon) (4 - 3\epsilon) (2 - 3\epsilon) \over 
      \epsilon^2 (2 - \epsilon)^2} \,
     {N_d^{-3} B_3(1,1,1,1)} \, \Gamma(1 - \epsilon/2)^{-3}
     \Gamma(1 + \epsilon/2)^{-3},
\end{eqnarray}
where $d = 4 -\epsilon$ and 
\begin{equation}
N_d \equiv \, {2\over (4\pi)^{d/2} \Gamma(d/2)}.
\end{equation}
Using the expansion of $B_4$ in powers of $\epsilon$ reported in 
\cite{Broadhurst_92}, we obtain 
\begin{equation}
B_3(1,1,1,1) = \, N_d^3\left\{
{2\over \epsilon^3} + {5\over 6 \epsilon^2} + 
{1\over 8\epsilon} (1 + 2 \pi^2) - {103\over96} + {5\pi^2\over 48} +
   O(\epsilon)\right\} .
\end{equation}
Let us now describe the algorithm for the other three cases.
Let us begin with $B_2(n_1,n_2,n_3)$. The basic relations are
\begin{eqnarray}
(d - 2 n_1 - n_3) B_2 &=& \left[-2 n_1\ {\bf 1}^+ - 
   n_3 \ {\bf 3^+} + n_3\ {\bf 3}^+ ({\bf 1}^- - 
   {\bf 2}^-) \right] B_2,
\label{relB2_1}
\\
(d-n_1-n_2-n_3) B_2 &=& 
   - (n_1\ {\bf 1}^+ + n_2\ {\bf 2}^+ + n_3\ {\bf 3}^+) B_2,
\label{relB2_2}
\end{eqnarray}
where ${\bf 1}^\pm B_2(n_1,n_2,n_3) = B_2(n_1\pm 1,n_2,n_3)$ and so on.
Using (\ref{relB2_1}) and the relation obtained interchanging 2 and 3 
we can reduce each $B_2(n_1,n_2,n_3)$ to integrals of the form 
$B_2(m_1,1,1)$ and to $B_2(m_1,m_2,0)$. The latter terms are the product of
two one-loop integrals. To deal with the former ones, 
sum to (\ref{relB2_1}) the relation obtained interchanging 2 and 3 in 
(\ref{relB2_1}). Then, use Eq. (\ref{relB2_2}) to eliminate the terms 
$(n_2\ {\bf 2}^+ + n_3\ {\bf 3}^+)B_2$ and 
$(n_2\ {\bf 2}^+ + n_3\ {\bf 3}^+){\bf 1}^- B_2$. The new relation can 
be used to express each $B_2(m_1,1,1)$ in terms of $B_2(1,1,1)$ and 
of the product of one-loop integrals. The integral 
$B_2(1,1,1)$ has been computed exactly in any dimension, 
obtaining \cite{Broadhurst-etal_93,Davydychev-etal_93}:
\begin{eqnarray}
B_2(1,1,1) &=& - {2 N^2_d \over \epsilon^2 (1-\epsilon)(2-\epsilon)} \,
    \Gamma(1 + \epsilon/2)^2\, \Gamma(2 - \epsilon/2)^2 
\nonumber \\
&& \times \left\{ 3^{1/2-\epsilon/2} 
   {\pi \Gamma(\epsilon)\over \Gamma(\epsilon/2)^2} + 
   {3\over2}(1-\epsilon)\, {}_2F_1(1,\epsilon/2;3/2;1/4)\right\}.
\end{eqnarray}
Expanding in powers of $\epsilon$ we obtain
\begin{equation}
B_2(1,1,1) = N_d^2\left\{ - {3\over 2\epsilon^2} - {3\over 4\epsilon}
 - {3\over 4} + {3\lambda\over4} - {\pi^2\over 8} +
 \epsilon \left[- {3\over4} - {\pi^2\over 16} + 
  {3\over 8}(\lambda + \gamma_E \lambda + Q_1)\right] + 
 O(\epsilon^2)\right\},
\end{equation}
where 
\begin{eqnarray}
\lambda &\equiv& {3\sqrt{\pi}\over 4} F_1 + 
     {\sqrt{3}\pi \over 6} (\gamma_E + \log 3) =\, 
     {2\over \sqrt{3}} {\rm Cl}_2\left({\pi\over 3}\right) =\,
     {1\over3} \psi'\left({1\over3}\right) - {2\pi^2\over 9},
\\
Q_1 &\equiv& {3\sqrt{\pi}\over 4} F_2 - 
     {\sqrt{3}\pi \over 12} (\gamma_E + \log 3)^2 - 
     {\sqrt{3}\pi^3\over 24},
\end{eqnarray}
$\psi(x)$ is the logarithmic derivative of the $\Gamma$-function and 
${\rm Cl}_2(x)$ is Clausen's polylogarithm \cite{Lewin_book}.
Here $F_1$ and $F_2$ are the following sums:
\begin{eqnarray}
F_1 &\equiv& {1\over 2} \sum_{n=0}^\infty 
   {\Gamma(n+1)\psi(n+1)\over \Gamma(n+3/2)} 4^{-n},
\\
F_2 &\equiv& {1\over 4} \sum_{n=0}^\infty
   {\Gamma(n+1)\over \Gamma(n+3/2)} 
   \left(\psi'(n+1) + \psi(n+1)^2\right) 4^{-n}.
\end{eqnarray}
Let us now discuss the integrals $G(n_1;n_2,n_3;n_4,n_5)$. 
First of all, note that if one of the indices is zero, $G$ 
can be written as a $B_3$ integral or as a product of a $B_2$ and of 
a one-loop integral. We will now show that any $G(n_1;n_2,n_3;n_4,n_5)$
can be expressed in terms of $G(1;1,1;1,1)$ and of $G$-integrals 
in which one of the indices is zero.
Using the integration-by-parts technique, we obtain the following 
relations: 
\begin{eqnarray}
&& 
 (d - 2 n_1 - n_3 - n_5) G =\, \left[
  - (2 n_1\ {\bf 1}^+ + n_3\ {\bf 3}^+ + n_5\ {\bf 5}^+) \right.
\nonumber \\ 
&& \qquad\qquad\qquad\qquad +
  \left. n_3\ {\bf 3}^+ ({\bf 1}^- - {\bf 2}^-) + 
  n_5\ {\bf 5}^+ ({\bf 1}^- - {\bf 4}^-) \right] G,
\label{relG_1}
\\
&& 
 (d - 2 n_4 - n_5) G =\, \left[
  - (2 n_4\ {\bf 4}^+ + n_5\ {\bf 5}^+) -
  n_5\ {\bf 5}^+ ({\bf 1}^- - {\bf 4}^-) \right] G,
\label{relG_2}
\\
&& 
 \left({3d\over2} - n_1 - n_2 - n_3 - n_4 - n_5 \right) G = \, 
 - (n_1\ {\bf 1}^+ + n_2\ {\bf 2}^+ + n_3\ {\bf 3}^+ + 
    n_4\ {\bf 4}^+ + n_5\ {\bf 5}^+ ) G,
\label{relG_3}
\\
&& 
 \left[n_5 - n_4 + n_4\ {\bf 4}^+ - n_5\ {\bf 5}^+ +
     n_4\ {\bf 4}^+ ({\bf 5}^- - {\bf 1}^-) - 
     n_5\ {\bf 5}^+ ({\bf 4}^- - {\bf 1}^-) \right] G =\,  0.
\label{relG_4}
\end{eqnarray}
First we consider Eq. (\ref{relG_1}) and sum to it the three relations 
obtained interchanging $(2\leftrightarrow 3)$, 
$(23\leftrightarrow 45)$, and $(23\leftrightarrow 54)$. Then, we use 
Eq. (\ref{relG_3}) eliminating $(n_2\ {\bf 2}^+ + n_3\ {\bf 3}^+ + 
    n_4\ {\bf 4}^+ + n_5\ {\bf 5}^+ ) G$ and 
$(n_2\ {\bf 2}^+ + n_3\ {\bf 3}^+ + n_4\ {\bf 4}^+ + n_5\ {\bf 5}^+ ) 
{\bf 1}^- G$. Replacing $n_1$ by $n_1 - 1$, we obtain a relation which
(for $n_1 \ge 2$)
expresses $G(n_1;n_2,n_3;n_4,n_5)$ in terms of 
$G(m_1;m_2,m_3;m_4,m_5)$ with $m_1 < n_1$. Therefore, in a finite number of 
steps we express the original integrals in terms of 
$G(1;n_2,n_3;n_4,n_5)$ and of integrals in which one of the indices is zero.
Then, we consider Eq. (\ref{relG_1}) and use the relation obtained 
previously to eliminate ${\bf 1}^+ G$. The new relation can be used to 
reduce $n_3$ to 1. Interchanging 2 and 3 we can similarly reduce 
$n_2$ to 1. We end up with $G(1;1,1;n_4,n_5)$ and with terms in which 
one index is zero. To further reduce the integrals we consider 
Eq. (\ref{relG_2}). Replacing $n_4$ by $n_4-1$ we obtain a relation 
which reduces each integral to the form $G(1;1,1;1,n_5)$. 
Finally, we consider (\ref{relG_4}) and use Eq. (\ref{relG_2})
to eliminate the terms ${\bf 4}^+G$ and ${\bf 4}^+{\bf 5}^- G$,
obtaining a relation that reduces $n_5$ to 1.  We should now compute 
$G(1;1,1;1,1)$. Using the integration-by-parts technique  we first 
obtain the exact relation 
\begin{equation}
G(1;1,1;1,2) = G_0(1;1,1;1,2) - {\epsilon\over2} M_0(1,1,1;1,1,1).
\end{equation}
The integral $G_0(n_1;n_2,n_3;n_4,n_5)$ is given by Eq. (\ref{defG}) 
replacing $\Delta(k_2)^{n_2}$ with $\Delta_0(k_2)^{n_2}$ and 
$M_0(n_1,n_2,n_3;n_4,n_5,n_6)$ by Eq. (\ref{defM}) replacing 
$\Delta(k_2-k_3)^{n_4}$ with $\Delta_0(k_2-k_3)^{n_4}$; 
$\Delta_0(k)$ is the massless propagator $1/k^2$. For our purposes 
we only need the divergent part of $M_0(1,1,1;1,1,1)$ (the next contribution is
reported in \cite{Broadhurst_99}):
\begin{equation}
M_0(1,1,1;1,1,1) = N^3_d\, {\zeta(3)\over 2\epsilon} [1 + O(\epsilon)].
\end{equation}
The integral $G_0(1;1,1;1,2)$ can be computed exactly using the 
Mellin-Barnes technique \cite{Davydychev_91_92,Boos-Davydychev_91}.
We obtain:
\begin{eqnarray}
&& G_0(1;1,1;1,2) = {\pi^3\over32} {N_d^3\over \sin^3 (\pi\epsilon/2)}
{(2-\epsilon)^2\over (1-\epsilon)} \,
  \left\{ \vphantom{{1\over2}}
  {}_3F_2(\epsilon/2,\epsilon,1;1-\epsilon/2,1/2+\epsilon/2;1/4) 
  \right.
\nonumber \\
&& \quad + {1\over2}(1-\epsilon)\, {}_2F_1(1,\epsilon/2;3/2;1/4) + 
 \sqrt{\pi}\, {\Gamma\left({1+\epsilon\over2}\right)\over 
             \Gamma\left({\epsilon\over2}\right)} \,
 3^{-1/2-\epsilon/2}\, 2^{\epsilon-1}\, - 1 
\nonumber \\
&& \quad \left. \hphantom{\times} - 2^{-\epsilon} 
  {\Gamma\left({3\epsilon\over2}\right)
   \Gamma\left({1+\epsilon\over2}\right)
   \Gamma\left(1 - {\epsilon\over2}\right) \over 
   \Gamma\left({\epsilon\over2}\right) 
   \Gamma\left({1\over2}+\epsilon\right) }\,
   {}_2F_1(\epsilon,3\epsilon/2;1/2+\epsilon;1/4) 
  \right\}.
\end{eqnarray}
Using these expressions, we obtain finally for $\epsilon\to 0$,
\begin{eqnarray}
G(1;1,1;1,1) &= & N_d^3\left\{
  - {1\over \epsilon^3} - {4\over 3\epsilon^2} + 
    {1\over \epsilon}\left(- {25\over 12} + {3\lambda\over2} - 
    {\pi^2\over 8}\right) \right.
\nonumber \\ 
&& 
\left. - {19\over 6} - {\pi^2\over 6} + 
    {3\over 4} (Q_1 + \gamma_E\lambda + 2 \lambda) - 
    3 Q_2 + {7\over4} \zeta(3) \right\} + O(\epsilon),
\end{eqnarray}
where
\begin{equation}
Q_2 \equiv\, {\sqrt{\pi}\over 4} 
\sum_{n=1}^\infty {\Gamma(n)^2\over \Gamma(n+1/2)} 
{4^{-n}\over n!} \left(\psi(n+1/2) - \psi(n+1) + 2\log 2 + 
   {2\over n}\right).
\end{equation}
Finally we consider $M(n_1,n_2,n_3;n_4,n_5,n_6)$. Again we will 
assume all indices to be positive: if one is zero, the integral 
can be rewritten as an integral of the $G$ family. The basic 
relations we need are
\begin{eqnarray}
&& 
(d - 2 n_6 - n_1 - n_5) M = 
\nonumber \\ 
&& \qquad \left[
 -(2 n_6\ {\bf 6}^+ + n_1\ {\bf 1}^+ + n_5\ {\bf 5}^+) + 
 n_5\ {\bf 5}^+ ({\bf 6}^- - {\bf 4}^-) +
 n_1\ {\bf 1}^+ ({\bf 6}^- - {\bf 2}^-)\right] M,
\label{relM_1}
\\
&& 
 \left[n_5 - n_3 - n_5\ {\bf 5}^+ + n_3\ {\bf 3}^+ \right.
\nonumber \\ 
&& \left. \qquad + 
    n_5\ {\bf 5}^+ ({\bf 1}^- - {\bf 3}^-) -
    n_3\ {\bf 3}^+ ({\bf 1}^- - {\bf 5}^-) + 
    n_4\ {\bf 4}^+ ({\bf 2}^- + {\bf 5}^- - {\bf 6}^- - {\bf 3}^-)\right] M 
    =\, 0\; .
\label{relM_2}
\end{eqnarray}
To compute this family of integrals we rewrite Eq. (\ref{relM_2}) as 
$n_3\ {\bf 3}^+ M = (n_5\ {\bf 5}^+ + \ldots)M$. Then, replacing 
$n_3$ by $n_3-1$, we obtain a relation that reduces each integral 
to $M(n_1,n_2,1;n_4,n_5,n_6)$, or to $M$'s in which one index 
is zero, and that, therefore, can be rewritten as $G$-integrals. 
Using the relations that are obtained replacing 
$(123456)\to(312645)$ and $(123456\to321654)$ in Eq. (\ref{relM_2}), 
we can similarly reduce $n_4$ and $n_5$ to 1. To further simplify 
the integrals, we consider Eq. (\ref{relM_2}) replacing 
$(123456\to231564)$ and rewrite it as 
$M = {\bf 1}^- (n_6\ {\bf 6}^+ + \ldots) M/(n_1-1)$. If we apply 
this relation to $M(n_1,n_2,1;1,1,n_6)$ we obtain three types 
of terms: 
(a) $M(m_1,m_2,1;1,2,m_6)$ with $m_1+m_2+m_6+1<n_1+n_2+n_6$;
(b) $M(m_1,m_2,1;1,1,m_6)$ with $m_1+m_2+m_6< n_1+n_2+n_6$;
(c) $M(m_1,m_2,1;1,1,m_6)$ with $m_1+m_2+m_6 = n_1+n_2+n_6$ and $m_1<n_1$. 
Terms of type (a) can be eliminated applying repeatedly the relation we used 
to reduce $n_5$, generating terms of type (b). Repeating the procedure,
we end up with integrals in which either one index is zero or 
$n_1$ is 1. An analogous procedure
can be used to reduce $n_2$ to 1. At the end of this reduction 
we should only consider $M(1,1,1;1,1,n_6)$. 
We now consider Eq. (\ref{relM_1}) and use the relations we considered
in the reduction of $n_5$ and $n_1$ to eliminate ${\bf 5}^+ M$ and 
${\bf 1}^+ M$. Then, replacing $n_6$ by $n_6-1$, we obtain 
$M = {\bf 6}^- (\ldots) M /(n_6 - 1)$. This relation generates 
integrals of type $G$ and ${\bf 1}^+{\bf 6}^- {\bf 6}^-M$,
${\bf 5}^+{\bf 6}^- {\bf 6}^-M$, and ${\bf 6}^-M$. The first two terms 
can be eliminated by applying repeatedly previous relations. 
At the end of the procedure any integral $M$ is expressed 
in terms of $M(1,1,1;1,1,1)$ and of integrals of type $G$. The former 
integral has been computed in \cite{Broadhurst_99} obtaining
in $d = 4 - \epsilon$, 
\begin{eqnarray}
M(1,1,1;1,1,1) = \, N_d^3 \left({1\over 2\epsilon}\zeta(3) + H + 
  O(\epsilon)\right),
\end{eqnarray}
where 
\begin{equation}
H = - {1\over2}\, {\rm Cl}_2^2\left({\pi\over3}\right) + 
  2\, {\rm Li}_4\left({1\over2}\right) - 
  {17\pi^4\over 720} - {\pi^2\over12} \log^2 2 + {1\over 12} \log^4 2,
\end{equation}
and ${\rm Cl}_2(x)$ and ${\rm Li}_4(x)$ are polylogarithms \cite{Lewin_book}.
A numerical determination of $H$ was given in Ref. \cite{P-V-98}, rewriting
\begin{equation}
H = - {\pi^4\over 80} + {3\over4} \zeta(3) \log{3\over2} + 3\, \widehat{H},
\end{equation}
where $\widehat{H}$ is the integral appearing in Eq. (B.19) of Ref. 
\cite{P-V-98}. Numerically $\widehat{H} \approx -0.4346277$, so that 
$H\approx -2.155953$, in agreement with the numerical results of 
\cite{Broadhurst_99}. Unfortunately, in Ref. \cite{P-V-98},
we used $\widehat{H} \approx -0.04346277$ --- a factor of ten smaller --- 
obtaining an incorrect estimate of $H$.

\begin{table}[tbp]
\caption{Three-dimensional estimates of $\bar{g}^{*}$
from an unconstrained analysis, ``unc", and constrained analyses in
various dimensions. For the analyses which use the
estimates in $d=2$
we report two errors: the first one gives the uncertainty of
the resummation of the series, the second one expresses the
change in the estimate when the two-dimensional result varies
within one error bar.
}
\label{epsilon3d}
\begin{tabular}{cr@{}lr@{}lr@{}lr@{}lr@{}lr@{}l}
\multicolumn{1}{c}{$N$}&
\multicolumn{2}{c}{unc}&
\multicolumn{2}{c}{$d=1$}&
\multicolumn{2}{c}{$d=0,1$}&
\multicolumn{2}{c}{$d=2$}&
\multicolumn{2}{c}{$d=1,2$}&
\multicolumn{2}{c}{$d=0,1,2$}\\
\hline
0 &   1&.38(7) &  1&.42(2) &   &      & 1&.407(18+2) &  1&.396(16+4) & &
           \\ 
1 &   1&.40(8) &  1&.44(3) &  1&.41(3) & 1&.424(22+0) &  1&.408(19+1) & 
     1&.408(12+1) \\ 
2 &  1&.39(7)  &  1&.42(3) &  1&.43(2) & 1&.426(19+6) &  1&.427(10+11) & 
     1&.425(8+16) \\ 
3 &  1&.39(7) &   1&.40(3) &  1&.41(2) & 1&.410(19+1) &   1&.420(7+2) & 
     1&.426(7+2) \\ 
4 &  1&.37(7) &   1&.38(3) &  1&.38(2) & 1&.384(21+5) &  1&.389(10+11) & 
     1&.393(5+16) \\ 
8 &  1&.31(5) &  1&.30(3) &  1&.30(2) & 1&.301(19+1) &  1&.304(9+1) & 
     1&.307(4+2) \\ 
16 &   1&.210(26) &  1&.203(17) &    1&.200(12) & 1&.202(12+0) &  1&.201(7+0) & 
     1&.202(4+0) \\ 
24 &   1&.160(17) &  1&.155(12) &  1&.151(9) & 1&.152(9+0) &   1&.150(5+0) & 
      1&.150(4+0) \\ 
32 &  1&.129(14) &  1&.125(9) &  1&.122(7) & 1&.122(7+0) &   1&.120(4+0) & 
     1&.119(3+0) \\ 
48 &  1&.091(10) &  1&.089(6) &  1&.087(4) & 1&.087(4+0) &  1&.085(3+0) & 
     1&.085(2+0) \\ 
\end{tabular}
\end{table}

\begin{table}[tbp]
\caption{Two-dimensional estimates of $\bar{g}^{*}$
obtained from analyses constrained at $d=1$ and at $d=0,1$.
}
\label{epsilon2d}
\begin{tabular}{cr@{}lr@{}l}
\multicolumn{1}{c}{$N$}&
\multicolumn{2}{c}{$d=1$}&
\multicolumn{2}{c}{$d=0,1$}\\
\hline
0 &  1&.72(4) &  &     \\ 
1 &  1&.84(6) &  1&.76(5) \\ 
2 &  1&.80(7) &  1&.82(3) \\ 
3 &  1&.73(8) &  1&.75(3) \\
4 &  1&.66(9) &  1&.67(4) \\ 
8 &  1&.46(6) &  1&.46(3) \\ 
16 &  1&.29(4) &  1&.28(2) \\ 
24 &  1&.21(3) &  1&.20(2) \\ 
32 &   1&.17(3) &   1&.16(1) \\ 
48 &  1&.12(2) & 1&.11(1) \\ 
\end{tabular}
\end{table}

\begin{table}[tbp]
\caption{Three-dimensional  estimates of $\bar{g}^*\equiv g^* (N+8)/(48\pi)$.
The two results of Ref.~\protect\cite{B-C-98} are relative to the 
cubic and bcc lattice respectively.  
}
\label{summarygr}
\begin{tabular}{cr@{}lr@{}lcr@{}l}
                   \multicolumn{1}{c}{$N$} 
                 & \multicolumn{2}{c}{$\epsilon$-exp. } 
                 & \multicolumn{2}{c}{$d=3$ $g$-exp. \cite{G-Z-98}} 
                 & \multicolumn{1}{c}{HT \cite{B-C-98}}
                 & \multicolumn{2}{c}{IHT} \\
\hline
0 & 1&.396(20) & 1&.413(6)  & 1.388(5),$\;$ 1.387(5) && \\
1 & 1&.408(13) & 1&.411(4)  & 1.408(7),$\;$ 1.407(6) & 1&.402(2) \cite{C-P-R-V-99} \\ 
2 & 1&.425(24) & 1&.403(3)  & 1.411(8),$\;$1.406(8)  & 1&.396(4) \cite{C-P-R-V-00} \\ 
3 & 1&.426(9)  & 1&.391(4)  & 1.409(10),$\;$1.406(8) &&\\
4 & 1&.393(21) & 1&.377(5)  & 1.392(10),$\;$ 1.394(10) &&\\ 
\end{tabular}
\end{table}

\begin{table}[tbp]
\caption{Three-dimensional estimates of $r_6$
for various values of $N$  from an unconstrained analysis of the
$\epsilon$-expansion and constrained analyses in $d=1$ and $d=0,1$.
The result in brackets for $N=0$ has been obtained using our conjectured
value for $r_6$  at $d=0$, see Eq.~(\protect\ref{d0res}).  
}
\label{r6d3}
\begin{tabular}{cccc}
\multicolumn{1}{c}{$N$}&
\multicolumn{1}{c}{unc}&
\multicolumn{1}{c}{$d=1$}&
\multicolumn{1}{c}{$d=0,1$}\\
\tableline \hline
0 &   2.180(80) &  2.148(22) & [2.146(15)] \\
1 &  2.077(69) &  2.057(31) & 2.065(18) \\
2 &  1.980(65) &  1.955(28) & 1.969(12) \\
3 &  1.889(63) &  1.859(21) & 1.867(9) \\
4 &  1.812(66) &  1.778(23) & 1.780(8) \\
8 &  1.580(78) &  1.546(25) & 1.537(15) \\ 
16 &  1.333(38) &  1.310(17) & 1.300(18) \\ 
32 &  1.125(13) &  1.117(4) & 1.110(9) \\
48 &  1.036(10)&  1.033(2) & 1.029(4) 
\end{tabular}
\end{table}

\begin{table}[tbp]
\caption{Two-dimensional estimates of $r_6$
for various values of $N$  from constrained analyses in $d=1$ and $d=0,1$.
The result in brackets for $N=0$ has been obtained using our conjectured
value at $d=0$.  
}
\label{r6d2}
\begin{tabular}{ccc}
\multicolumn{1}{c}{$N$}&
\multicolumn{1}{c}{$d=1$}&
\multicolumn{1}{c}{$d=0,1$}\\
\tableline \hline
0 &   3.745(47) &  [3.740(23)] \\ 
1 &  3.671(68) &  3.691(28) \\ 
2 &  3.494(58) &  3.530(18) \\ 
3 &  3.308(41) &  3.328(12) \\ 
4 &  3.155(44) &  3.159(12) \\ 
8 &  2.747(45) &  2.721(19) \\ 
16 &  2.368(34) &  2.335(24) \\ 
32 &  2.074(10) &  2.052(13) \\ 
48 &  1.950(4) &  1.937(6) \\ 
\end{tabular}
\end{table}

\begin{table}[tbp]
\caption{Three-dimensional estimates of $r_8$ 
for various values of $N$  from an unconstrained analysis of the
$\epsilon$-expansion and constrained analyses in $d=1$ and $d=0,1$.
The result in brackets for $N=0$ has been obtained using our conjectured
value for $r_8$ at $d=0$.  
}
\label{r8d3}
\begin{tabular}{cccc}
\multicolumn{1}{c}{$N$}&
\multicolumn{1}{c}{unc}&
\multicolumn{1}{c}{$d=1$}&
\multicolumn{1}{c}{$d=0,1$}\\
\tableline \hline
0 &  0.1(2.3) &  2.19(1.16) & [3.13(53)] \\ 
1 & $-$0.4(1.9) &  1.76(80) & 2.75(39) \\
2 & $-$0.8(1.7) &  0.75(75) & 2.08(45) \\
3 & $-$1.2(1.5) &  0.01(48) & 0.97(28) \\
4 & $-$1.4(1.1) & $-$0.50(31) & 0.19(23) \\
8 & $-$1.8(4) & $-$1.39(18) & $-$1.18(9) \\
16 & $-$1.7(4) & $-$1.57(7) &  $-$1.54(4) \\
32 & $-$1.3(1) & $-$1.23(3) & $-$1.24(2) \\
48 & $-$0.97(5) & $-$0.96(1) & $-$0.969(4) 
\end{tabular}
\end{table}

\begin{table}[tbp]
\caption{Two-dimensional estimates of $r_8$
for various values of $N$  from constrained analyses in $d=1$ and $d=0,1$.
The result in brackets for $N=0$ has been obtained using our conjectured
value for $r_8$ at $d=0$.  
}
\label{r8d2}
\begin{tabular}{ccc}
\multicolumn{1}{c}{$N$}&
\multicolumn{1}{c}{$d=1$}&
\multicolumn{1}{c}{$d=0,1$}\\
\tableline \hline
0 &  24.6(2.3) & [27.0(9)] \\ 
1 &  23.8(1.3) &  26.5(5) \\ 
2 &  19.7(1.4) & 23.2(6) \\ 
3 &  16.1(1.0) & 18.8(4) \\ 
4 &  13.5(8) &  15.4(3) \\ 
8 &  8.1(4)  &  8.7(2) \\ 
16 &  5.0(2) &  5.1(1) \\ 
32 &  3.87(5) &  3.82(2) \\ 
48 &  3.71(2) &  3.64(1) 
\end{tabular}
\end{table}

\begin{table}
\caption{
Three-dimensional estimates of $r_{6}$ and $r_8$.
When the original reference reports only estimates of $g_{2j}$ 
the errors we quote for $r_{2j}$  have been calculated
by considering the estimates of $g_{2j}$ as uncorrelated. 
}
\label{summaryd3}
\begin{tabular}{ccr@{}lr@{}lr@{}lr@{}l}
\multicolumn{1}{c}{}&
\multicolumn{1}{c}{}&
\multicolumn{2}{c}{$\epsilon$-exp.}&
\multicolumn{2}{c}{$d=3\;\;g$-exp.}&
\multicolumn{2}{c}{HT}&
\multicolumn{2}{c}{ERG}\\
\tableline \hline
$N=0$&$r_6$&  2&.148(22) [this work]  &   2&.11(9) [this work] &   &&    \\ 
 &     &  2&.1(3)~\cite{P-V-98-b} &&   &&   &&    \\ \hline
 &$r_8$&   2&(1) [this work]&&   &&   &&    \\ 
 &     &  6&(5)~\cite{P-V-98-b} &&   &&   &&    \\ \hline\hline

$N=1$&$r_6$&2&.058(11)~\cite{P-V-98-b} & 2&.053(8) \cite{G-Z-98,G-Z-97}&2&.048(5) \cite{C-P-R-V-99}& 1&.92 \cite{T-W-94}\\ 
 &     &  2&.12(12) \cite{G-Z-98,G-Z-97}   & 2&.060 \cite{S-O-U-K-99} & 1&.99(6) \cite{B-C-97} &  & \\ 
 &     &   &                       &       & & 2&.157(18) \cite{Z-L-F-96} &   &   \\ \hline

 &$r_8$&  2&.48(28) \cite{P-V-98-b}  & 2&.47(25) \cite{G-Z-98,G-Z-97}  & 2&.28(8) \cite{C-P-R-V-99} & 2&.18 \cite{T-W-94}\\ 
 &     &  2&.42(30)\cite{G-Z-98,G-Z-97}     & 2&.496 \cite{S-O-U-K-99} & 2&.7(4) \cite{B-C-97} &  &    \\\hline

 &$r_{10}$& $-$20&(15) \cite{P-V-98-b}& $-$25&(18) \cite{G-Z-98,G-Z-97}&$-$13&(4) \cite{C-P-R-V-99} & & \\ 
 &     &  $-$12&.0(1.1) \cite{G-Z-98,G-Z-97} & &  & $-$10&(2) \cite{C-P-R-V-99} &  &   \\
 &     &   & & &  & $-$4&(2) \cite{B-C-97} &  &   \\\hline \hline

$N=2$&$r_6$&  1&.969(12) [this work]              & 1&.967 \cite{S-O-U-K-99}& 1&.951(14) \cite{C-P-R-V-00}& 1&.83 \cite{T-W-94}\\ 
 &     &  1&.94(11)  \cite{P-V-98-b}  &  1&.970(40) [this work]& 2&.2(6) \cite{Reisz-95} &  &    \\\hline

 &$r_8$&   2&.1(0.9)   [this work]            & 1&.641 \cite{S-O-U-K-99}& 1&.36(9) \cite{C-P-R-V-00}& 1&.4 \cite{T-W-94} \\ 
 &     &  3&.5(1.3)  \cite{P-V-98-b}  &  & & & &  &    \\\hline\hline

$N=3$&$r_6$&   1&.867(9) [this work]  & 1&.880 \cite{S-O-U-K-99}& 2&.1(6) \cite{Reisz-95}& 1&.74 \cite{T-W-94} \\ 
 &     &  1&.84(9)  \cite{P-V-98-b}  &  1&.884(32) [this work] &  & &  &    \\\hline

 &$r_8$&   1&.0(0.6) [this work]               & 0&.975 \cite{S-O-U-K-99}&  & & 0&.84 \cite{T-W-94} \\ 
 &     &  2&.1(1.0)  \cite{P-V-98-b}  &  & & & &  &    \\\hline\hline

$N=4$&$r_6$&   1&.780(8) [this work] & 1&.803 \cite{S-O-U-K-99}& 1&.9(6) \cite{Reisz-95}& 1&.65 \cite{T-W-94} \\ 
 &     &  1&.75(7)  \cite{P-V-98-b}  &  1&.809(27) [this work]&  & &  &   \\\hline

 &$r_8$&   0&.2(0.4) [this work]      & 0&.456 \cite{S-O-U-K-99}&  & & 0&.33 \cite{T-W-94} \\ 
 &     &  1&.2(1.0)  \cite{P-V-98-b}  &  & & & &  &    \\
\end{tabular}
\end{table}

\end{document}